\documentclass[a4paper,oneside,font=normalsize]{article}
\pdfoutput=1 

\usepackage[textwidth=17cm,textheight=24cm]{geometry}
\usepackage{hyperref}
\usepackage{authblk}
\usepackage[pdftex]{graphicx}
\usepackage[font=small]{caption}   

\usepackage{placeins}

\providecommand{\keywords}[1]{\textbf{\textit{Keywords---}} #1}

\def\jpsi{\mbox{$J\!/\!\psi$}}%

\begin{document}

This is an author-created, un-copyedited version of an article accepted for publication in JINST. 
IOP Publishing Ltd is not responsible for any errors or omissions 
in this version of the manuscript or any version derived from it. 
The Version of Record is available online at \url{https://doi.org/10.1088/1748-0221/12/07/c07006}.

\title{\boldmath The PANDA DIRC Detectors at FAIR}


\author[a,1]{C.~Schwarz\footnote{Corresponding author.},} 
\author[a,b]{A.~Ali,}
\author[a]{A.~Belias,} 
\author[a]{R.~Dzhygadlo,} 
\author[a]{A.~Gerhardt,} 
\author[a]{K.~G\"{o}tzen,} 
\author[a,2]{G.~Kalicy\footnote{present address: The Catholic University of America, Washington, USA},}
\author[a,b]{M.~Krebs,}
\author[a]{D.~Lehmann,} 
\author[a,b]{F.~Nerling,} 
\author[a,3]{M.~Patsyuk\footnote{present address: Massachusetts Institute of Technology, Cambridge, USA},}
\author[a,b]{K.~Peters,}
\author[a]{G.~Schepers,} 
\author[a]{L.~Schmitt,}
\author[a]{J.~Schwiening,} 
\author[a]{M.~Traxler,}
\author[a]{M.~Z\"{u}hlsdorf,}
\author[c]{M.~B\"{o}hm,}
\author[c]{A.~Britting,}
\author[c]{W.~Eyrich,}
\author[c]{A.~Lehmann,}
\author[c]{M.~Pfaffinger,}
\author[c]{F.~Uhlig,}
\author[d]{M.~D\"{u}ren,}
\author[d]{E.~Etzelm\"{u}ller,}
\author[d]{K.~F\"{o}hl,}
\author[d]{A.~Hayrapetyan,}
\author[d]{K.~Kreutzfeld,}
\author[d]{B.~Kr\"{o}ck,}
\author[d]{O.~Merle,}
\author[d]{J.~Rieke,}
\author[d]{M.~Schmidt,}
\author[d]{T.~Wasem,}
\author[e]{P.~Achenbach,}
\author[e]{M.~Cardinali,}
\author[e]{M.~Hoek,}
\author[e]{W.~Lauth,}
\author[e]{S.~Schlimme,}
\author[e]{C.~Sfienti,}
\author[e]{M.~Thiel,}
\author[f]{L.~Allison} 
\author[f]{C.~Hyde}
\affil[a]{GSI Helmholtzzentrum f\"ur Schwerionenforschung GmbH, Darmstadt, Germany}
\affil[b]{Goethe University, Frankfurt a.M., Germany}
\affil[c]{Friedrich Alexander-University of Erlangen-Nuremberg, Erlangen, Germany}
\affil[d]{II. Physikalisches Institut, Justus Liebig-University of Giessen, Giessen, Germany}
\affil[e]{Institut f\"{u}r Kernphysik, Johannes Gutenberg-University of Mainz, Mainz, Germany}
\affil[f]{Old Dominion University, Norfork, U.S.A.}

\maketitle
\flushbottom

\abstract{
The PANDA detector at the international accelerator 
Facility for Antiproton and Ion Research in Europe (FAIR) 
addresses fundamental questions of hadron physics. 
An excellent hadronic particle identification (PID) will be accomplished 
by two DIRC (Detection of Internally Reflected Cherenkov light) counters
in the target spectrometer.
The design for the barrel region covering polar angles between \mbox{$22^\circ$} to \mbox{$140^\circ$} 
is based on the successful BABAR DIRC 
with several key
improvements, such as fast photon timing and a compact imaging region.
The novel Endcap Disc DIRC will cover the smaller forward angles 
between \mbox{$5^\circ$} (\mbox{$10^\circ$}) to \mbox{$22^\circ$} in the vertical (horizontal) direction.
Both DIRC counters  will use lifetime-enhanced
microchannel plate PMTs for photon detection in combination with fast readout electronics. 
Geant4 simulations and tests with several prototypes at various beam facilities have
been used to evaluate the designs and validate the expected
PID performance of both PANDA DIRC counters.
}

\vspace{10mm}


\section{Particle Identification of PANDA}
\label{sec:intro}

\begin{figure}[t]
\captionsetup{width=0.8\textwidth}
\centering 
\includegraphics[width=.6\textwidth]{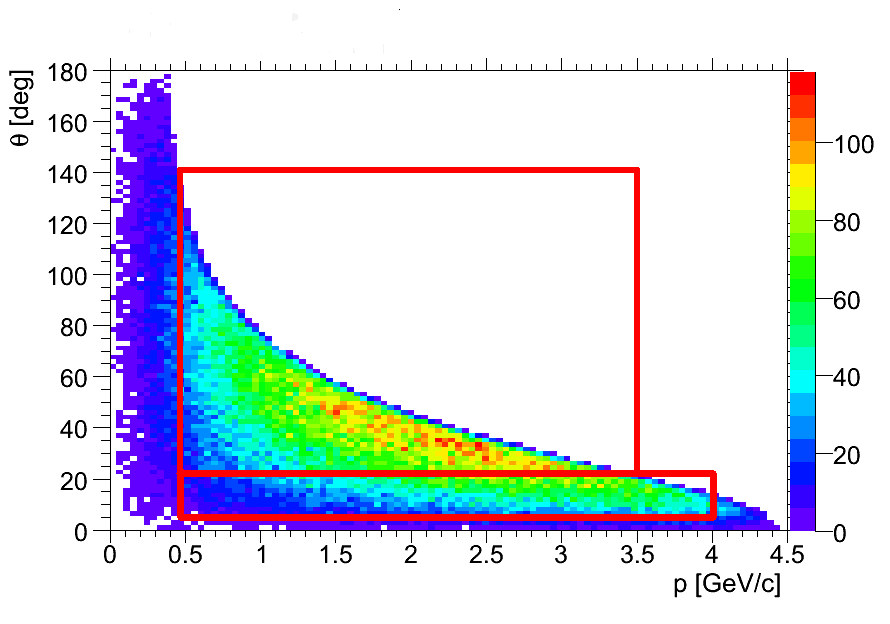}
\caption{\label{fig:acceptance} Intensity of kaons from 
\mbox{$\jpsi \rightarrow K^+K^-\gamma$}~ from simulated antiproton proton annihilations 
at \mbox{$\sqrt{s}=3.1$~GeV/$c$} as function of polar angle $\theta$ 
and momentum p. The top and bottom rectangles denote the acceptance of the Barrel DIRC and 
the Endcap Disc DIRC, respectively.}
\end{figure}
The PANDA experiment~\cite{panda1} will be one of the four flagship experiments at the new 
international accelerator complex FAIR (Facility for Antiproton and Ion Research) in Darmstadt, Germany.
PANDA will perform unique experiments using the high-quality antiproton beam with momenta in the 
range of 1.5~GeV/$c$ to 15~GeV/$c$, stored in the HESR (High Energy Storage Ring) 
to explore fundamental questions of hadron physics in the charmed and multi-strange hadron sector 
and deliver decisive contributions to the open questions of QCD \cite{panda-physics}.
The cooled antiproton beam colliding with a fixed proton or nuclear target will allow hadron production 
and formation experiments. 
Two complementary operating modes are planned,
named high luminosity and high resolution. The
high luminosity mode with a momentum resolution
of \mbox{$\Delta$p/p = 10$^{-4}$} and stochastic cooling will have a luminosity of
$2\cdot10^{32}cm^{-2}s^{-1}$. For the high resolution mode
 \mbox{$\Delta$p/p = 4$\cdot10^{-5}$} will be achieved with electron cooling 
for momenta up to p = 8.9 GeV/c. The cycle-averaged luminosity is expected to be
$10^{31}cm^{-2}s^{-1}$.
Excellent Particle Identification (PID) is crucial to the success of the PANDA physics program.
The PID system comprises a range of detectors 
using different technologies.
Dedicated PID devices, such as several Time-of-Flight and Cherenkov counters and a Muon 
detection system, are combined with PID information delivered by the 
Micro Vertex Detector and the Straw Tube Tracker  
as well as by the Electromagnetic Calorimeter.
The DIRC concept was introduced and successfully used by the BaBar 
experiment~\cite{adam2005,coyle2004} 
where it provided excellent $\pi$/K separation up to 4.2~GeV/$c$ and proved to 
be robust and easy to operate.
The PANDA Barrel DIRC, modeled after the BaBar DIRC, will surround the interaction point at a 
distance of about 50~cm and cover the central region of polar angles $22^\circ < \theta < 140^\circ$
while the novel Endcap Disc DIRC will cover the smaller forward angles, 
$5^\circ < \theta < 22^\circ$ and $10^\circ < \theta < 22^\circ$ in the vertical and horizontal 
direction, respectively. As shown in Fig.~\ref{fig:acceptance} the Barrel DIRC needs to 
separate pions from kaons for momenta up to
\mbox{$3.5$~GeV/$c$} with a separation power of at least 3 standard deviations (s.d.) while the Endcap 
Disc DIRC aims for a PID up to particle momenta of \mbox{$4$~GeV/$c$} with at least 4 s.d. separation. 

\section{Barrel DIRC}

%
%
\begin{figure}[t]
\captionsetup{width=0.8\textwidth}
\centering
\includegraphics[width=0.48\columnwidth]{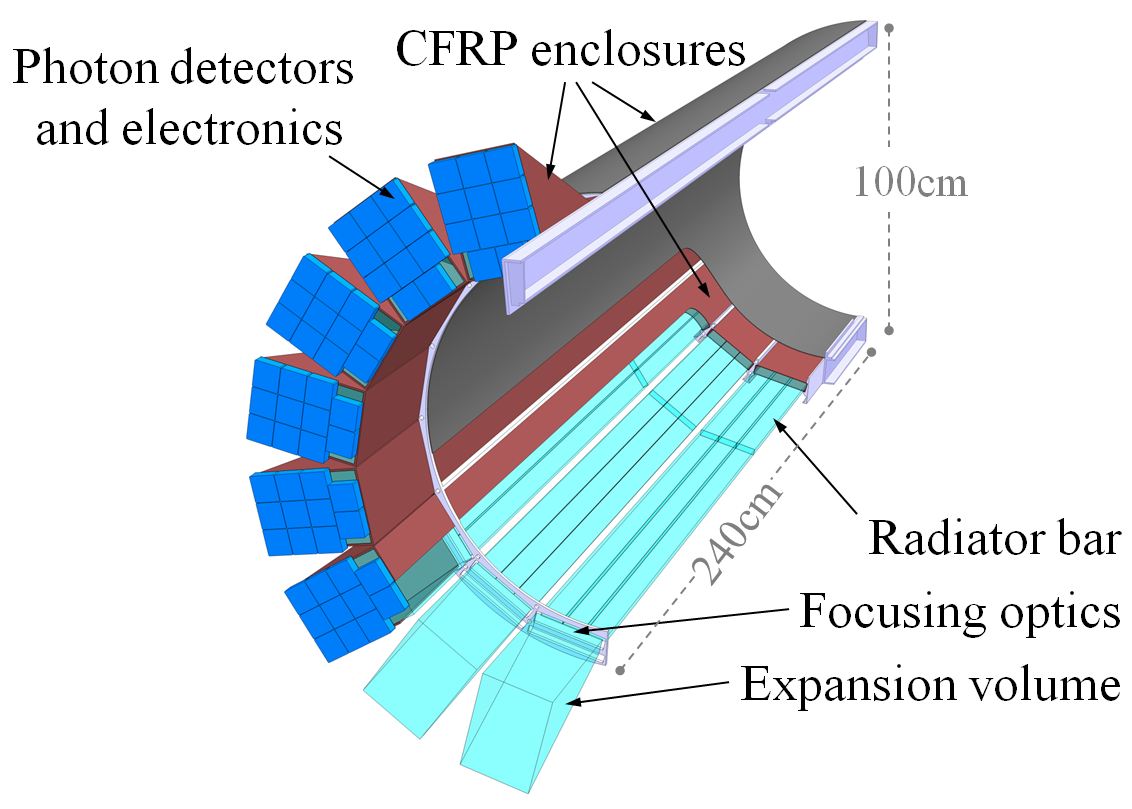}
\includegraphics[width=0.48\columnwidth]{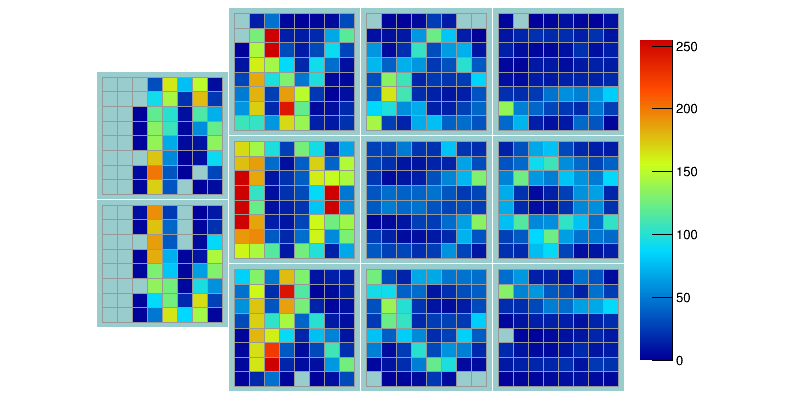}
\caption{\label{fig:barrel-dirc-design}
Left: Schematic of the Barrel DIRC baseline design. Only one half of the detector is shown.
Right: Geant simulation of the baseline geometry of the PANDA Barrel DIRC.
The colored histogram shows the accumulated hit pattern from 1000~$K^{+}$
at 3.5~GeV/$c$ momentum and $55^{\circ}$ polar angle.
}
\end{figure}
\begin{figure}[b]
\captionsetup{width=0.8\textwidth}
\centering 
\includegraphics[width=.49\textwidth]{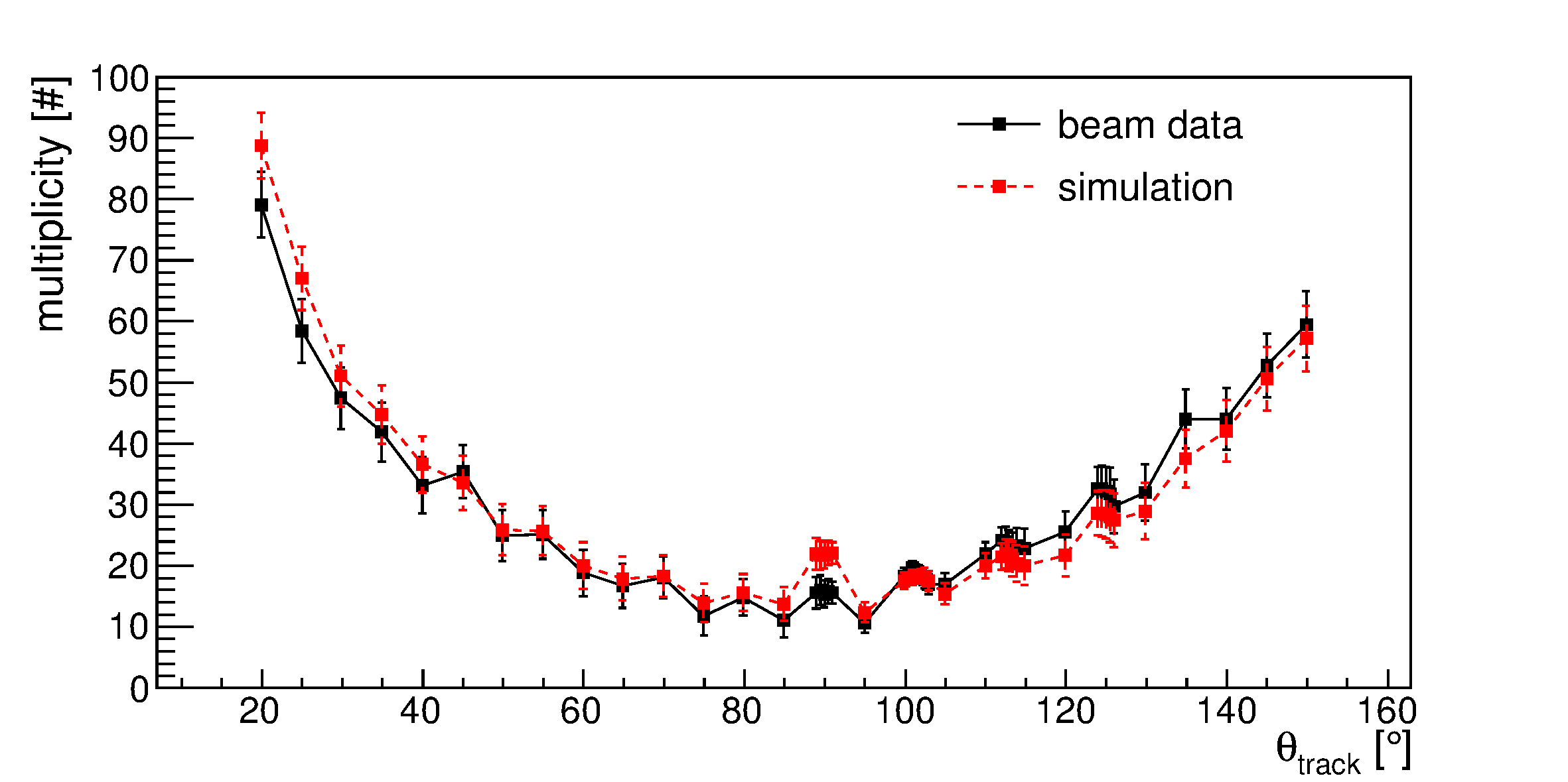}
\includegraphics[width=.49\textwidth]{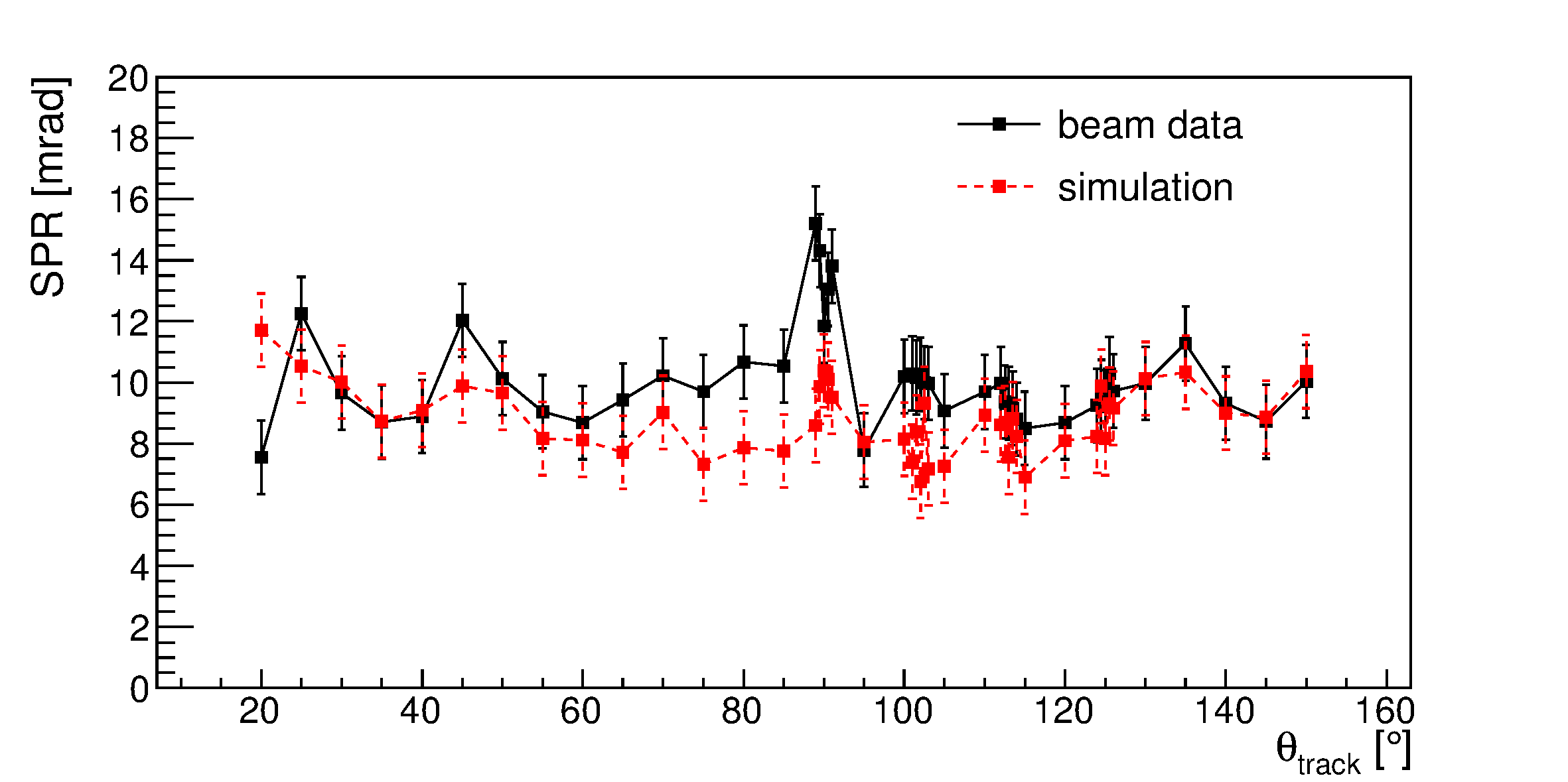}
\caption{\label{fig:nph-spr} 
Photon yield (left) and SPR (right)
as a function of the track polar angle for the narrow
bar and the 3-layer spherical lens for tagged protons at
7 GeV/$c$ beam momentum in data (black) and Geant
simulation (red). The error bars correspond to the RMS
of the distribution in each bin.
The use of lower-quality, older sensors affected predominantly the angles 
around $90^\circ$ (left and right).
}
\end{figure}
Since the space limits for the PANDA Barrel DIRC are tight,
several design modifications 
were required compared to the BaBar DIRC.
Due to the optical and mechanical spe\-ci\-fi\-cations the fabrication 
of the radiator bars remains one of the dominant cost drivers for DIRC counters.
A significant cost reduction is only possible if fewer pieces have to be polished.
Detailed physical simulation studies demonstrated that reducing the number of
bars per bar box from 5  (32~mm width) to 3 (53~mm width) does not
affect the PID performance since the lens system is able to correct for 
the increase in bar size. 
The overall design of the PANDA experiment required that the large water
tank used by the BaBar DIRC is replaced by a compact expansion volume (EV),
placed inside the detector.
Fused silica as material and separated smaller units as expansion volume 
were already favored by the SuperB FDIRC \cite{dey2015}  and the Belle~II TOP \cite{krizan2014}.
The baseline design of the PANDA Barrel DIRC detector \cite{dzhygadlo2016} is
shown in Fig.~\ref{fig:barrel-dirc-design} (left).
Sixteen optically isolated sectors, each comprising a bar box and a solid 
fused silica prism, surround the beam line in a 16-sided polygonal barrel 
with a radius of 476~mm and cover the polar angle range of  
$22^\circ < \theta < 140^\circ$. 
Each bar box contains three bars of 17~mm thickness, 53~mm width, 
and 2400~mm length, 
placed side-by-side, separated by a small air gap.
A flat mirror is attached to the forward end of each bar to reflect 
photons towards the read-out end, where they are focused by a 3-component 
spherical compound lens on the back of a $30$~cm-deep solid prism, 
made of synthetic fused silica. 
The location and arrival time of the photons are measured by an array 
of 11 lifetime-enhanced Microchannel-plate (MCP) PMTs \cite{lehmann2016} 
with a spatial and timing precision of about 2~mm and $100$~ps, 
respectively. The MCP-PMTs are read out by an updated version of the HADES trigger and 
readout board (TRB)~\cite{trb3-jinst} in combination with a front-end
amplification and discrimination card mounted directly on the 
MCP-PMTs~\cite{cardinali:padiwa}.
%
A detailed physical simulation of the PANDA Barrel DIRC was developed 
in Geant4 \cite{geant4} and an accumulated hit pattern from these simulations 
is shown in Fig.~\ref{fig:barrel-dirc-design} (right). 

The goal of the test beam campaign at the CERN PS in 2015 and 2016 was the validation of the
PID performance of the baseline design and of the wide plate.
The prototype comprised the essential 
elements of 
one PANDA Barrel DIRC sector:
A narrow fused silica bar (17.1 $\times$ 35.9 $\times$ 1200.0~mm$^3$) 
or a wide fused silica plate (17.1 $\times$ 174.8 $\times$ 1224.9~mm$^3$), 
coupled on one end to a flat mirror, on the other end to a focusing lens, 
and the fused silica prism as EV (with a depth of 300~mm and a top angle of 45$^\circ{}$).
A very fast time-of-flight (TOF) system~\cite{tof-jinst-pe}, positioned directly in 
the beam, was used for $\pi/p$ tagging. 
The reconstructed photon yield as a function of the track polar angle is shown 
in Fig.~\ref{fig:nph-spr}\,(left) for the configuration with the narrow bar radiator and 
the 3-layer spherical lens.
The number of Cherenkov photons from the beam data (black) ranges from 12 to 80 
and is in agreement with simulations (red). 
The single photon Cherenkov angle resolution for the same data set 
is shown in Fig.~\ref{fig:nph-spr}\,(right). 
The beam data and simulation are consistent within the RMS of the distributions 
for the forward and backward angles.
Most of the data were taken with the beam momentum
of 7 GeV/$c$. The $\pi/p$ Cherenkov angle difference
at this momentum (8.1 mrad) is close to the $\pi/K$
Cherenkov angle difference at 3.5 GeV/$c$ (8.5 mrad).
The design with the narrow bar and the spherical lens is found to meet or exceed
the PID requirements for PANDA. It is robust against timing deterioration and delivers 
excellent $\pi/K$ separation for the imaging reconstruction methods. 

The prototype tests demonstrated that the figures of merit and the $\pi$/K separation power of
the geometry based on narrow bars exceeded the
PANDA PID requirements for the entire pion and
kaon phase space. 
After improving several key aspects of the prototype configuration in 2016, 
the observed $\pi$/p separation power for the wide plate is $2.8^{+0.4}_{-0.2}$
standard deviations (s.d.) without focusing. 
With the 2-layer cylindrical lens
the $\pi$/p separation is $3.1^{+0.1}_{-0.1}$ s.d., in good agreement with the prototype simulation, which predicts
a $3.3^{+0.1}_{-0.1}$ s.d. separation value. However, the design with
narrow bars provides a larger margin for error and 
can be expected to perform significantly better.
\begin{figure}[tp]
\captionsetup{width=0.8\textwidth}
\centering
\includegraphics[width=0.9\columnwidth]{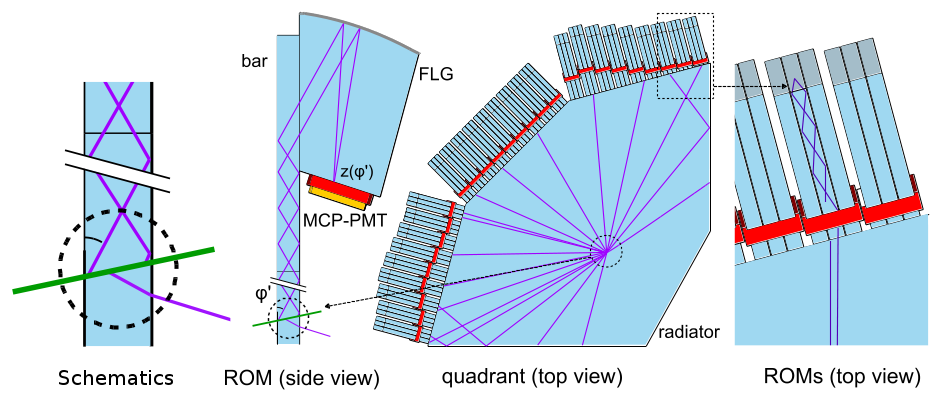}
\caption{\label{fig:edd-design}
Schematics: A charged particle traverses the radiator plate and emits Cherenkov photons 
(violet lines) at certain angles. Most of them undergo total 
internal reflection and are thus trapped inside the plate. ROM side view: The photons are 
guided by a rectangular bar (prism) and focused by a focussing element (FEL) with a 
cylindrical mirror. They are registered by an MCP-PMT that measures the position of the 
photon within a wavelength range that is defined by a filter. Quadrant, top view and ROM, 
top view: The azimuthal angle is determined from the position of the prism where the photon 
is registered.
}
\end{figure}

\section{Endcap Disc DIRC}

\begin{figure}[t]
\captionsetup{width=0.8\textwidth}
\centering
\includegraphics[width=0.36\columnwidth]{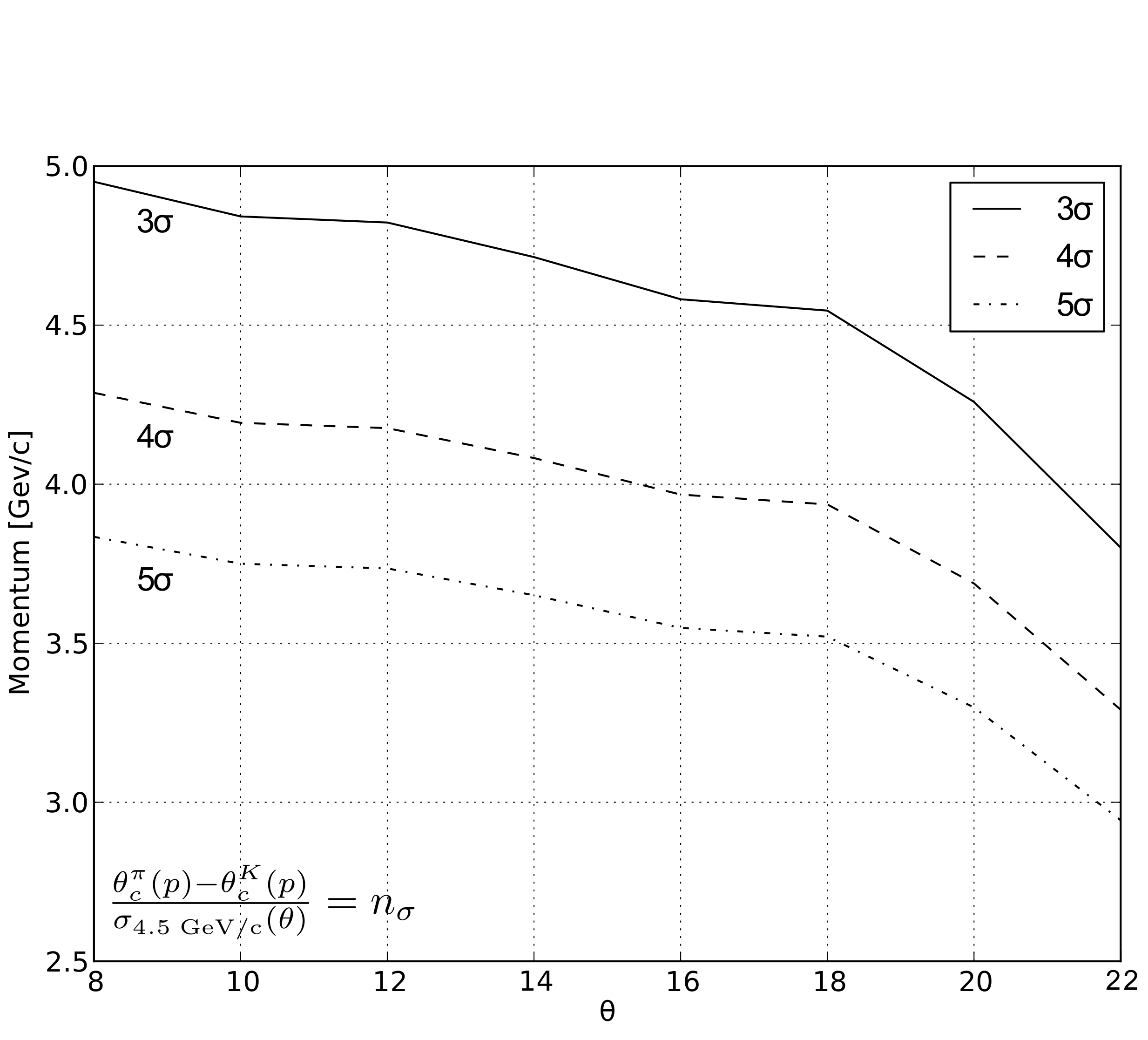}
\qquad
\includegraphics[width=0.43\columnwidth]{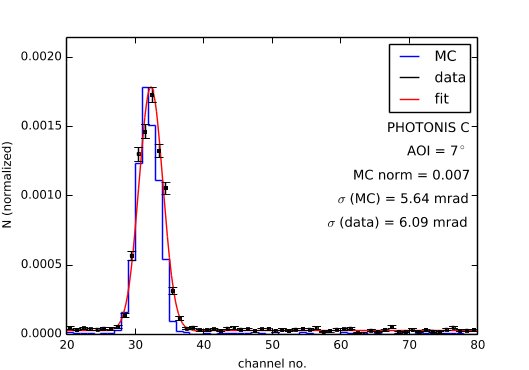}
\caption{\label{fig:edd_sep_power}
Left:Pion/kaon separation computed from the
distribution of reconstructed Cherenkov angles for a
ROM consisting of a $2"$ tube with 0.5 mm pixel width.
Right: Comparison of data and Monte-Carlo of the single-photon resolutions at 10 GeV/$c$ beam momentum.
}
\end{figure}

The forward region of the target spectrometer will be equipped with a Endcap Disc DIRC and it will be the first time that this type of
detector will be used in a high-performance $4\pi$ experiment \cite{dueren2016}. 
The detector will be divided into four independent quadrants which form a disc with a diameter of 
about 2 m and an active area of roughly 3.5~m$^2$. Each quadrant consists of
one 2 cm thick radiator with 27 Read-Out Modules (ROMs) attached to its outer sides (see 
Fig. \ref{fig:edd-design}). One ROM combines three Focusing Elements (FEL) with one MCP-PMT and the
corresponding readout electronics. Current MCP-PMT options have a segmented anode of \mbox{$6 \times 128$}
pixels (Hamamatsu) or \mbox{$3 \times 100$} pixels (Photonis). 
The curved side of the FEL is coated with a reflective aluminum 
surface and optimized to match the position resolution of the MCP-PMTs. 
An optical longpass or bandpass filter is foreseen to reduce
the chromatic error and to limit the number of photons for a longer 
lifetime of the photocathode of the MCP-PMT. For every charged track crossing 2 cm fused silica 22 
detected photons are expected within a wavelength range of \mbox{$365-400 nm$}.
The ToFPET ASIC \cite{rolo} is foreseen for the readout of the MCP-PMTs . 
Geant4 simulation results \cite{geant4} 
of the expected separation power are shown in Fig. \ref{fig:edd_sep_power}.
The goal of a PID with at least 4 s.d. separation up to particle momenta of \mbox{$4$~GeV/$c$} is 
achievable for polar angles up to \mbox{$16^\circ$}. For larger angles, the charged track is too close 
to the rim of the disc deteriorating the angular resolution. A larger disc could remedy 
this effect, but would not fit into the PANDA detector due to spatial requirements. 
A prototype test was done at the T9 beam line of PS East Area at CERN in 2015 together with the Barrel DIRC.
The polished radiator was a 
\mbox{$500 \times 500 \times 20~$mm$^3$} plate made of NIFS-S by Nikon. The optical system was 
completed by three FEL/prism pairs which had
been coupled to the radiator.
A figure of merit for the performance of the Endcap Disc DIRC is the single photon resolution which can
be measured with a single FEL. 
Fig. \ref{fig:edd_sep_power} shows the accumulated photon 
hits for a fixed angle of the hadron beam with respect to
the DIRC radiator at 10 GeV/$c$. 
The x-axis represents the channel or strip numbers of the MCP-PMT anode. 
One channel corresponds to 3.5 mrad.
The beam test demonstrated the good agreement of the simulation with the experimental data.
%
%

\section{Conclusion}

The beam test with the PANDA Barrel DIRC prototype at CERN in 2015 and 2016 successfully
validated the PID performance of both radiator geometries, the narrow bar with the spherical lens,
and the wide plate with the cylindrical lens. 
The PANDA Barrel DIRC design with
narrow bars provides a larger margin for error and
can be expected to perform significantly better during
the first PANDA physics run due to the dependence
of the wide plate geometry on excellent timing.
Due to these key performance advantages, the geometry
with the narrow bars and the 3-layer spherical
lens was selected as the baseline design for the
PANDA Barrel DIRC.
After initial tests of component prototypes (optics, photon detectors, readout) 
of the Endcap Disc DIRC in the lab demonstrated the expected performance,   
a large system prototype was used in 2015 at the T9 beam line at CERN to measure the 
single photon resolution.
The radiator and FELs were made of high quality fused silica, with a design and 
quality comparable to the final one. 
The reconstruction of Cherenkov photons and the single photon 
resolution of the prototype agree well with Monte Carlo expectations.

\section*{Acknowledgments}

This work was supported by 
HGS-HIRe, 
HIC 
for FAIR, 
BNL eRD14 and 
U.S. National Science
Foundation PHY-125782.
We thank the GSI, DESY and CERN staff for the opportunity to use 
the beam facilities and for their on-site support.

\end{document}